\documentclass[conference]{IEEEtran}
\IEEEoverridecommandlockouts
\usepackage{cite}
\usepackage{amsmath,amssymb,amsfonts}
\usepackage{graphicx}
\usepackage{textcomp}
\usepackage{xcolor}
\usepackage{algorithm}
\usepackage{algpseudocode}
\usepackage{makecell}
\usepackage{multirow}
\usepackage{subcaption}
\usepackage{url}

\def\BibTeX{{\rm B\kern-.05em{\sc i\kern-.025em b}\kern-.08em
    T\kern-.1667em\lower.7ex\hbox{E}\kern-.125emX}}

\newcommand{\RNum}[1]{\uppercase\expandafter{\romannumeral #1\relax}}

\newtheorem{definition}{Definition}
\begin{document}

\title{EasyTUS: A Comprehensive Framework for Fast and Accurate Table Union Search across Data Lakes}

\author{\IEEEauthorblockN{Tim Otto}
\IEEEauthorblockA{\textit{Heterogeneous Information Systems} \\
\textit{RPTU Kaiserslautern-Landau}\\
Kaiserslautern, Germany \\
tim.otto@cs.rptu.de}
}

\maketitle

\begin{abstract}

Data lakes enable easy maintenance of heterogeneous data in its native form. While this flexibility can accelerate data ingestion, it shifts the complexity of data preparation and query processing to data discovery tasks. One such task is Table Union Search (TUS), which identifies tables that can be unioned with a given input table.
In this work, we present EasyTUS, a comprehensive framework that leverages Large Language Models (LLMs) to perform efficient and scalable Table Union Search across data lakes. EasyTUS implements the search pipeline as three modular steps: Table Serialization for consistent formatting and sampling, Table Representation that utilizes LLMs to generate embeddings, and Vector Search that leverages approximate nearest neighbor indexing for semantic matching.
To enable reproducible and systematic evaluation, in this paper, we also introduce TUSBench, a novel standardized benchmarking environment within the EasyTUS framework. TUSBench supports unified comparisons across approaches and data lakes, promoting transparency and progress in the field.
Our experiments using TUSBench show that EasyTUS consistently outperforms most of the state-of-the-art approaches, achieving improvements in average of up to 34.3\% in Mean Average Precision (MAP), up to 79.2× speedup in data preparation, and up to 7.7× faster query processing performance. Furthermore, EasyTUS maintains strong performance even in metadata-absent settings, highlighting its robustness and adaptability across data lakes.

\end{abstract}


\section{Introduction}
The exponential growth of heterogeneous data generation across the domains calls for innovative strategies for storing and leveraging information at scale. The data may appear in structured, semi-structured, or unstructured formats, and require minimal processing and quick gateway access \cite{DBLP:journals/tkde/HaiKQJ23}. Data lakes have emerged as a practical solution aligned with big data principles, enabling easy storage of vast volumes of raw, uncurated data, where metadata may be lacking or ambiguous. While this flexibility accelerates data ingestion by several folds, it shifts the burden of data discovery and preparation to query execution time. The primary challenge in querying from data lakes is the difficulty of post-hoc identifying relevant or valuable data \cite{DBLP:journals/pvldb/NargesianZMPA19}. Among the key techniques for querying structured data in such environments is Table Union Search (TUS) \cite{DBLP:journals/pvldb/NargesianZPM18}. TUS aims to identify other tables within and across multiple data lakes that can be merged with a given query table via a union operation. This capability is valuable not only for data analysts seeking to enrich their information pool, but also for machine learning pipelines, where augmenting datasets with unionable tables can improve model generalization by increasing the number of data points. 

\begin{figure}
    \centering
    \includegraphics[width=\linewidth]{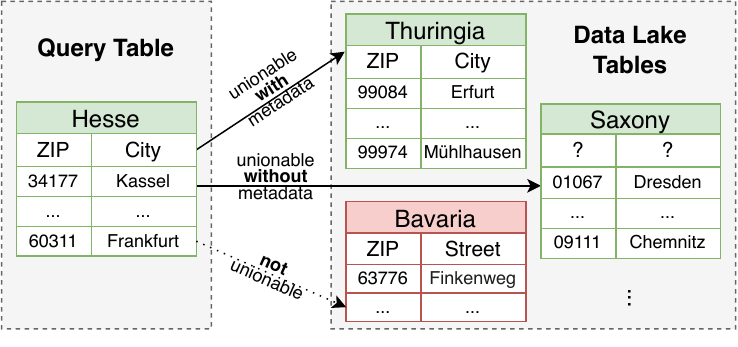}
    \caption{Illustration of Table Union Search on a single data lake. The query table (left) and potential unionable tables from the data lake (right) are highlighted in green. Tables that are not unionable are marked in red.}
    \label{fig:example}
\end{figure}

A simple example of a TUS query executed within a single data lake is illustrated in Figure~\ref{fig:example}. In this scenario, the query table contains ZIP codes and city names from the German state of Hesse. The goal is to find other tables containing ZIP codes and city names from different regions of Germany that can be unioned with the query table. Tables such as Thuringia and Saxony are valid unionable results, as their schemas match and their tuples can be seamlessly appended to the query table. However, table Bavaria contains ZIP codes and street names instead of city names. Although it may be joinable based on the shared attribute ZIP code, it is not unionable because of differing column semantics. Performing a union operation in this case would result in an ill-defined outer union, where non-matching columns would be padded with null values, therefore, defeating the purpose of a clean union operation.

Earlier Table Union Search approaches \cite{DBLP:conf/icde/BogatuFP020, DBLP:journals/pacmmod/KhatiwadaFSCGMR23} have explored techniques such as metadata analysis, column-level set similarity, and ontology-based annotation. However, their accuracy is constrained by the limitations of data lakes: metadata is often sparse or inaccurate, exact value matches are rare, and domain-specific ontologies, particularly for numerical or non-English data, are frequently unavailable. In addition, querying across multiple data lakes is not supported. Beyond the challenge of capturing semantic relationships between queries and targets (e.g. ensuring that a query about German cities ranks tables containing other German cities higher than those with cities from different countries), computationally expensive operations such as discovering functional dependencies become non-trivial as data volumes grow.

Recent approaches \cite{DBLP:journals/pvldb/FanWLZM23, DBLP:conf/acl/HuWQLSFKKWY23, DBLP:journals/corr/abs-2407-01619} rely on extensively fine-tuning the models and manually configuring the hyperparameters for embedding-based comparisons, aiming to improve semantic matching. However, these approaches also have challenges: smaller models often fail to generalize across domains or languages, and fine-tuning models on specific data lakes restricts applicability, making searches across multiple data lakes impossible. Given that data lakes can expand rapidly, generalization across data lakes and domains, and runtime efficiency, spanning data preparation (offline time) and query execution (online time), are critical. In such environments, it is impractical to re-train and fine-tune TUS approaches for each set of new data lakes.

This paper introduces \textit{EasyTUS}, a novel Table Union Search framework that leverages Large Language Models (LLMs) without relying on metadata, ontologies, or fine-tuning. Trained on broad and diverse corpora, EasyTUS offers strong generalizability across domains and languages, eliminating the need for time-consuming, task-specific adaptation. Its framework is built on LLMs and operates directly on native table content, requiring only data lakes at offline time and a query table as input during online time, with no additional parameter configuration. This makes it especially effective for poorly curated data lakes. Importantly, removing fine-tuning steps unlocks the capability to query across multiple data lakes.

Alongside EasyTUS, this paper introduces \textit{TUSBench}, a novel benchmarking environment for a robust comparison method with reproducible results, evaluating both accuracy and runtime efficiency of table union approaches across different data lakes. TUSBench is integrated within the EasyTUS framework and specifically designed to assess the effectiveness and efficiency of Table Union Search approaches and data lakes in a standardized and streamlined manner while underlining the importance of reproducible benchmarking results. It supports one-command execution, enforces consistent input/output formats, and facilitates result sharing and publication, ultimately promoting transparency, reproducibility, and accelerating progress in the field.
Our contributions include:

\begin{enumerate}
    \item We propose \textit{EasyTUS}, a comprehensive framework that leverages LLMs for fast and accurate Table Union Search across multiple data lakes. EasyTUS operates independently of metadata, ontologies, or fine-tuning, relying solely on the underlying data in the data lakes.
    \item EasyTUS also includes \textit{TUSBench}, a novel benchmarking environment that enables the reproducible comparison of the state-of-the-art and future Table Union Search approaches across various data lakes. TUSBench also supports varying the configurations of individual approaches, e.g. evaluating different models for any specific LLM-based TUS approach. It unifies input/output formats, streamlines evaluation, and facilitates the publication of reproducible results.
    \item We provide a detailed analysis of the existing state-of-art Table Union Search techniques.
    \item In the experiments, we demonstrate that EasyTUS outperforms the state-of-the-art in most cases in terms of accuracy and runtime efficiency.
\end{enumerate}

In the next section, we formally define the Table Union Search problem. Section~\ref{sec:easytus} introduces the architecture of EasyTUS, followed by Section~\ref{sec:related_work}, which discusses related work and contrasts existing approaches with EasyTUS, motivating the need for a standardized benchmarking environment. Section~\ref{sec:tusbench} presents TUSBench, our proposed benchmarking environment, after which we report experimental results in Section~\ref{sec:experiments} and conclude.

The source code for EasyTUS and TUSBench is publicly available at: \url{https://sci-git.cs.rptu.de/t_otto18/easytus}

\section{Background}
Formally, Table Union Search \cite{DBLP:journals/pvldb/NargesianZPM18} can be defined as a top-$k$ retrieval problem.

\begin{definition}
Given a set of tables in a data lake $\mathcal{T}$ and a query table $q$, the \textit{Table Union Search @$k$} task retrieves a subset $\mathcal{T}_q \subseteq \mathcal{T}$ with $|\mathcal{T}_q| = k$, such that each union $q \cup t_i$, with $t_i \in \mathcal{T}_q$, is well-defined and
\[
score(q \cup t_i) \geq score(q \cup t_{i+1}),
\]
where $score(\cdot)$ measures the union compatibility between a target table in the data lake and the query table.
\end{definition}

Among the well-defined union operations between two tables $t_1$ und $t_2$, with respective columns $t_1 = (c_1, \dots, c_n)$ and $t_2 = (c_1, \dots, c_m)$, two special cases can be highlighted that frequently occur in real-world data lakes:

\begin{itemize}
    \item \textit{One-to-One}: $|t_1|=|t_2|$. The union operation is a bijective function that maps each unique column in $t_1$ to a unique column in $t_2$.
    \item \textit{Onto}: $|t_1| > |t_2|$ (resp. $|t_2| > |t_1|$). The union operation is an injective function that maps all columns of $t_2$ (resp. $t_1$) to unique columns in $t_1$ (resp. $t_2$).
\end{itemize}

It is important to note that an \textit{onto} mapping may semantically correspond to a \textit{one-to-one} mapping. This can occur, for instance, when a single column in $t_1$ is split into two separate columns in $t_2$, preserving the overall information content despite the structural mismatch. To complete the classification, an ill-defined union operation is a partial injective function that maps a subset of columns $t_1' \subset t_1$ to a subset $t_2' \subset t_2$.

\section{EasyTUS Architecture}
\label{sec:easytus}
\begin{figure*}
    \centering
    \includegraphics[width=\linewidth]{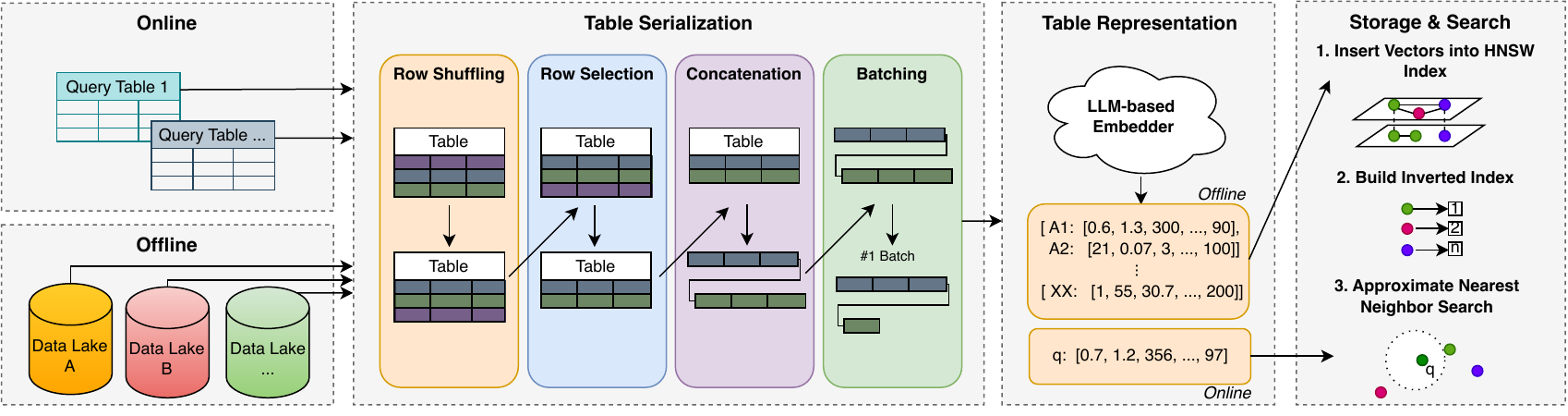}
    \caption{Overview of the EasyTUS architecture for Table Union Search across multiple data lakes. The core steps, Table Serialization and Table Representation, are executed in both the offline and online phases. In the offline phase, embedding vectors for all tables from individual data lakes are generated and persisted, while in the online phase, an approximate nearest neighbor search is performed using the query table vector and the persisted embeddings.}
    \label{fig:system}
\end{figure*}
EasyTUS is designed on the principle of architectural simplicity, incorporating only the minimal necessary functionality while improving both accuracy and time efficiency. The framework is structured into multiple steps, namely Table Serialization, Table Representation, Storage, and Search, as illustrated in Figure~\ref{fig:system}. A distinction is made between offline and online time. Offline time refers to executing all EasyTUS pipeline steps, with the generated embedding vectors for tables across data lakes persisted in the vector database. Online time corresponds to executing only the first two pipeline steps followed by the search, applied when a query table is received. In this case, embedding vectors for the query table are generated identically but are never stored.

\subsection{Table Serialization}
Using Large Language Models for Table Union Search requires representing tables from data lakes as serialized inputs, i.e., as serialized tokens of table values, to the LLMs. In general, two strategies can be distinguished: \textit{zero-shot} and \textit{few-shot} \cite{DBLP:journals/corr/abs-2402-07927}. In the \textit{few-shot} strategy, metadata such as column names or table context is included alongside table values, while \textit{zero-shot} relies solely on the table values itself. Since metadata in data lakes can be ambiguous, missing, or unreliable, EasyTUS employs the zero-shot approach to ensure robustness and general applicability. Regardless of the specific LLM used, there is typically a maximum input token limit, which gives rise to two fundamental challenges: determining how many table values can be included in a single serialization without exceeding the model’s token limit, and deciding which parts of the table (i.e., which columns) should be selected to ensure an informative yet complete representation. Addressing these challenges requires effective strategies.

\subsubsection{Batching Strategy}
In general, each LLM has a maximum token limit $N$ that constrains the length of a single table serialization. Additionally, usually their respective APIs support \textit{batch processing}, where multiple serializations can be ingested in parallel, subject to a global batch limit $M$. In cases where batch processing is not supported, we set $N = M$ without loss of generality. The core constraint for batching is defined as:

\begin{equation}
    \sum_{0...b,0...i} \text{tokens}[b][i] \leq b \cdot M \,\land\, \forall_{0...b}\sum_{0...i} \text{tokens}[b][i] \leq N
\end{equation}

where $\text{tokens}[b][i]$ denotes the number of tokens in the $i$-th serialization of batch $b$. The goal is to maximize the number of serializations in each batch to reduce the number of API calls and overall execution time. EasyTUS applies a greedy batching algorithm, as shown in Algorithm~\ref{alg:batch}, that selects just enough rows per table to fit within the table serialization limit $N$, and then packs as many such serializations into a batch until the batch-wide maximum token limit $M$ is reached.

\begin{algorithm}
\caption{EasyTUS Greedy Batching Algorithm}
\label{alg:batch}
\begin{algorithmic}
\Require Table set $\mathcal{T}$, serialization token limit $N$, batch token limit $M$
\Procedure{Batching}{$\mathcal{T}$, $N$, $M$}
\State $Batches \gets [~]$, $Batch \gets [~]$, $Tokens \gets 0$
\ForAll{table $T \in \mathcal{T}$}
    \State $Rows \gets$ \Call{RowSelection}{$T$, x}
    \State $S \gets$ \Call{Serialize}{$Rows$}
    \State $D \gets$ \Call{Min}{0, $|S| - N$}
    \State $S \gets$ Remove last $D$ tokens from $S$
    \If{$Tokens + |S| > M$}
        \State Append $Batch$ to $Batches$
        \State $Batch \gets [~]$, $Tokens \gets 0$
    \EndIf
    
    \State Append $S$ to $Batch$
    \State $Tokens \gets Tokens + |S|$
\EndFor
\State Append $Batch$ to $Batches$
\EndProcedure
\end{algorithmic}
\end{algorithm}

\subsubsection{Selection Strategy}
The maximum token limit $N$ also poses the challenge of determining which columns to prioritize for the serialization in order to make the most effective use of the available token limit space. Prior work \cite{DBLP:journals/pvldb/PaulsenGD23} on finding most valuable columns commonly applies TF-IDF scoring across tables during offline time to identify the most informative columns. However, such global relevance scoring is computationally expensive and not easily maintainable in dynamic or frequently expanded data lakes.

Therefore, EasyTUS employs a pairwise-independent, lightweight row and column selection strategy. From each table, $x$ random rows are sampled, ignoring any metadata such as table or column names. The use of randomly selected rows reduces the risk of using approach-tailored data lakes that fail to reflect real-world scenarios, thereby improving generalizability. Within each row, column values are converted to strings and concatenated using a separator token \texttt{[SEP]} (e.g., a comma). Rows themselves are separated by a \texttt{[NEW]} token (e.g., \textbackslash n). If the resulting prompt exceeds the token limit $N$, tokens are removed from the end until the constraint is satisfied. If an entire table contains fewer data than the maximum token limit, every row from the table is included exactly once. In the case of a single row exceeding the token limit, its values are first shuffled before the algorithm proceeds with removing tokens from the end. The final serialized format looks like:
\[
\{\texttt{ID: "} v_{i,0}\; \texttt{[SEP]} \; v_{i,1}\; \texttt{[SEP]} \; \dots \; \texttt{[NEW]} \; v_{i+1,0} \; \dots\texttt{"}\}
\]

where \texttt{ID} represents the composite key of data lake and table name, and $v_{i,j}$ denotes the value of the $j$-th column in the $i$-th randomly selected row.

By reducing the number of rows, EasyTUS preserves the maximal number of columns within each serialization. This trade-off favors breadth over depth, increasing the likelihood that relevant column semantics are captured in the embedding despite tight token constraints. Importantly, this method generates a semantic representation for the entire table rather than for individual columns. This reduces the risk of overestimating unionability due to superficial similarities in isolated columns, such as primary keys with common integer patterns, by considering the table's content as a whole. It also reduces computational costs by requiring only one representation per table, instead of computing and matching separate representations for each column. Furthermore, the LLM can capture structural relationships between columns based on their co-occurrence within the same serialization, thereby enhancing the overall semantic representation.

\subsection{Table Representation}
To generate vector representations of the table serializations across all batches, EasyTUS employs LLM-based embedders where each serialization in the batch is transformed into one embedding, resulting in a batch of corresponding vectors. This enables querying based on semantic similarity, thus allowing for more flexible, nuanced, and efficient comparison across diverse data lakes. EasyTUS is compatible with any embedding model: for example, models by OpenAI \cite{korade2024strengthening} or published on HuggingFace \cite{huggingface_2023}, demonstrating the flexibility and balance between cloud-based, high-performance models and on-premise solutions that prioritize data privacy. Within the EasyTUS framework, users only need to specify the model name and its token limit for the batching algorithm at offline time. It is essential to select a sufficiently large model to ensure robust cross-domain generalization. EasyTUS makes no use of fine-tuning or post-processing of embeddings, prioritizing maximum throughput, scalability, and minimal configuration.

\subsection{Storage}
The embedding vectors are stored in a vector database. To support efficient retrieval, two types of indexes are created: a Hierarchical Navigable Small World (HNSW) \cite{DBLP:journals/pami/MalkovY20} index for fast approximate nearest neighbor (ANN) search over the embedding space, and an inverted index that maps embedding vectors back to their corresponding tables and data lakes.

\subsection{Table Union Search}
To ensure consistent and effective table retrieval across data lakes, the same Table Serialization and Representation methods that are applied to the data lake tables during offline time must also be applied to query tables during online time. Upon receiving a query table, the identical pipeline steps are executed, but the resulting embedding vector is never persisted. Instead, the online phase additionally performs an approximate nearest neighbor search using the query vector against the embedding vectors in the vector database.
EasyTUS supports any distance function, with cosine similarity as the default metric for similarity search. Alternative metrics, such as dot product or Euclidean distance, can also be used interchangeably. 

Given a query table $q$, the system returns a ranked list of $k$ embedding vectors with the highest similarity scores to $q$, enabling efficient retrieval of unionable tables across data lakes. Using the inverted index, these embedding vectors are then mapped back to the corresponding tables, producing a uniformly scored, ranked list of unionable tables spanning multiple data lakes.

\subsection{Incremental Table Additions}
Since data lakes are typically append-only and not actively maintained, challenges related to updates and deletions are largely avoided, i.e. new data gets added, but existing entries remain unchanged. As a result, the key operational concern is efficient table insertions, which must be fast and independent of the existing tables of the data lakes, thus making the total end-to-end execution time, comprising both offline and online time, a crucial metric. It also implies that all algorithms used for data selection or sampling must be stateless, meaning they cannot rely on tables that have already been processed. EasyTUS addresses this by using uniform embedding vectors: when a new table is added, its embedding can be computed independently and inserted directly into the index, without requiring re-processing of tables across the existing data lakes. Due to the use of random sampling, when rows in a table are added, updated, or deleted, it still have little to no effect on the generated embedding. Our sampling strategy offsets the table updates effectively. Only extensive modifications would noticeably affect the sampling; an assumption contrary to the intended use of data lakes.

\begin{table*}[!th]
    \centering
    \caption{Classification of Table Union Search Approaches}
    \begin{tabular}{|c|c|c|c|c|c|c|} \hline
         & \textbf{EasyTUS} & \textbf{D3L} \cite{DBLP:conf/icde/BogatuFP020} & \textbf{Santos} \cite{DBLP:journals/pacmmod/KhatiwadaFSCGMR23} & \textbf{Starmie} \cite{DBLP:journals/pvldb/FanWLZM23} & \textbf{AutoTUS} \cite{DBLP:conf/acl/HuWQLSFKKWY23} & \textbf{TabSketchFM} \cite{DBLP:journals/corr/abs-2407-01619} \\ \hline
        \textbf{Search Composition} & Individual & \makecell{Composite \\ (5 Indexes)} & \makecell{Composite \\ (KB + synth. KB)} & Individual & Individual & Pipelined \\ \hline
        \textbf{\makecell{Representation \\ Method}} & Transformer-based & Feature-based & Graph-based & Transformer-based & Transformer-based & Transformer-based \\ \hline
        \textbf{Learning Strategy} & No Learning & - & - & Self-Supervised & Self-Supervised & (Self-) + Supervised\\ \hline
        \textbf{Cross-Compitibility} & Yes & Partial & Partial & No & No & No \\ \hline
        \textbf{\makecell{Representation \\ Level}} & Table & Column & (Inter-) + Column & Column & Column & Table \\ \hline
        \textbf{Representation} & Embedding & Feature & Annotation & Embedding & Embedding & Embedding \\ \hline
        \textbf{\makecell{Data Type \\ Dependency}} & No & Yes & Yes & Yes & No & Yes \\ \hline
        \textbf{\makecell{Knowledge \\ Integration}} & No & No & External KB & No & No & No \\ \hline
        \textbf{Indexing Strategy} & HNSW & LSH-Forest & Inverted Index & HNSW & - & - \\ \hline
    \end{tabular}
    \label{tab:related_work}
\end{table*}

\section{Related Work}
\label{sec:related_work}
Existing Table Union Search approaches can be classified along several dimensions. An overview of how each method aligns with these criteria is summarized in Table~\ref{tab:related_work}. 

\textit{Search Composition}.
Approaches can either use a single technique (individual) or combine multiple techniques. For instance, D3L \cite{DBLP:conf/icde/BogatuFP020} uses a composite strategy with five different indexes, improving robustness by aggregating information from multiple perspectives. In contrast, TabSketchFM \cite{DBLP:journals/corr/abs-2407-01619} applies a pipelined method where the output of one step forms the input of the next step.

\textit{Representation Method}.
This refers to how table content is encoded for comparison. While many approaches, including EasyTUS and Starmie \cite{DBLP:journals/pvldb/FanWLZM23}, use transformer-based embeddings for semantic understanding, Santos \cite{DBLP:journals/pacmmod/KhatiwadaFSCGMR23} stands out by using graph-based representations, relying on structural and relational information derived from knowledge bases.

\textit{Learning Strategy}.
Learning strategies vary in supervision level. Starmie, for example, applies a self-supervised approach, learning from internal table relationships without labeled data. In contrast, TabSketchFM includes both self-supervised and supervised training to boost in-domain accuracy.

\textit{Cross-Compatibility}.
Some systems are designed to operate across multiple data lakes only in part, meaning individual components may function across lakes, but the complete system does not work in composition without reconfiguration. EasyTUS is the only approach supporting full cross-data lake compatibility, using a universal scoring function without requiring fine-tuning or knowledge base generation.

\textit{Representation Level}.
Table content can be represented at different granularities. Santos uses a mix of inter-column and column-level representations to capture both individual column semantics and their relational structure, enhancing expressive power for complex tables.

\textit{Representation}.
Information may be encoded as learned embeddings, extracted features, or external annotations. 

\textit{Data Type Dependency}.
Some approaches factor in column data types to improve matching precision. D3L is type-dependent, leveraging type-aware indexes to better align numeric and textual columns. In contrast, EasyTUS remains type-agnostic, treating all columns uniformly.

\textit{Knowledge Integration}.
Only Santos includes external domain knowledge, enriching its representation with semantic annotations from sources like YAGO \cite{DBLP:conf/www/SuchanekKW07}. 

\textit{Indexing Strategy}.
Efficient search at scale requires indexing. EasyTUS and Starmie both use HNSW, a scalable vector-based method, while Santos relies on an inverted index, suitable for textual matches derived from knowledge bases.

EasyTUS stands out by being the only training-free, fully cross-compatible method with minimal setup and strong scalability. Its transformer-based embeddings, table-level abstraction, and HNSW indexing enable fast, accurate search even across diverse and growing data lakes.
By contrast, each competing approach outlines their respective unique strengths. D3L emphasizes robustness via composite indexing and type-awareness. Santos enriches semantic matching with external knowledge. Starmie and AutoTUS \cite{DBLP:conf/acl/HuWQLSFKKWY23} provide a lightweight self-supervised alternative, and TabSketchFM improves expressiveness with task-specific fine-tuning.

These inherent differences across the state-of-the-art highlight the importance of a standardized benchmarking environment, which can enable consistent, reproducible evaluation across such diverse approaches.

\section{Novel Benchmarking Environment}
\label{sec:tusbench}
\begin{figure*}
    \centering
    \includegraphics[width=\linewidth]{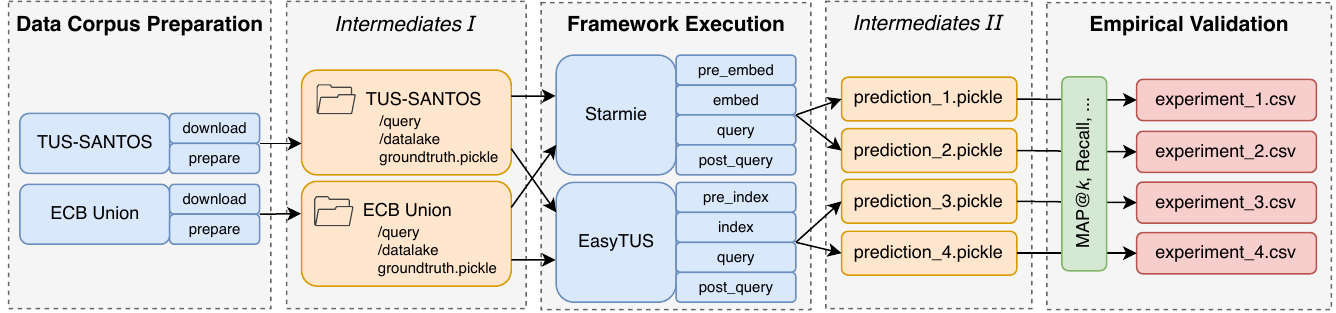}
    \caption{Overview of the TUSBench environment setup, illustrated with two example data lakes, namely TUS-SANTOS and ECB Union, and two example approaches, Starmie and EasyTUS. Blue boxes represent provided components, orange boxes indicate generated intermediates, green boxes denote customizable metrics, and red boxes highlight shareable results.}
    \label{fig:tusbench}
\end{figure*}
The diversity of data lakes and Table Union Search approaches leads to inconsistent methods for setting hyperparameters and measuring accuracy, typically through precision and recall. To ensure verifiable and reproducible results, there is a pressing need for a standardized framework that enables consistent evaluation of both approaches and data lakes, using common metrics for effectiveness and efficiency. Such standardization is essential for reproducible research and to foster progress in the scientific community. While a few benchmarking tools have been proposed \cite{DBLP:journals/pvldb/DengCCYCYSWLCJZJZWYWT24, DBLP:journals/corr/abs-2505-21329}, they often suffer from issues such as non-executability, faulty code, non-standardization, and irreproducible results. 

TUSBench addresses these challenges by structuring the execution into modular components and intermediate interfaces, ensuring a consistent and extensible workflow. The environment is structured into a pipeline consisting of three main stages: Data Corpus Preparation, Framework Execution, and Empirical Validation, which are connected by two well-defined intermediate interfaces (Intermediates \RNum{1} and Intermediates \RNum{2}). These standardized interfaces ensure consistency, extensibility, and reproducibly testing across all stages. An example TUSBench environment is shown in Figure~\ref{fig:tusbench}.

\subsubsection{Data Corpus Preparation}
TUSBench is compatible with any data lake, provided the data can be transformed into a tabular representation. This process is enforced through two sequential steps, executed once per data lake: \textit{download} and \textit{prepare}. The download step ensures that the executable code for respective approaches and the tables from the data lakes are stored separately, for example, in a GitHub repository and a storage server respectively. Since data lakes can occupy substantial storage space, this separation enables a dynamic and space-efficient environment where the data stays within the data lakes. It also allows new approaches to be shared without the need to transfer the data. The prepare step translates known file formats, structures, and locations into the standardized format required by the \textit{Intermediates \RNum{1}} interface, enabling seamless integration with any supported approach. If these steps have already been completed, they are skipped by default unless explicitly forced to re-run.

\subsubsection{Intermediates \RNum{1}}
The \textit{Intermediates \RNum{1}} interface defines a structure that enables consistent handoff between the Data Corpus Preparation and Framework Execution stage, ensuring interoperability across varying interpretations of data lake structures and approach inputs.

After completing the Data Corpus Preparation stage, the tables of each data lake are consolidated into a representational directory structure:

\begin{itemize}
    \item \texttt{/query} – Contains query tables
    \item \texttt{/datalake} – Contains (non-)unionable tables
    \item \texttt{groundtruth.pickle} – Maps query tables to corresponding unionable tables
\end{itemize}

This standardized layout enables any approach to interact with any data lake avoiding the need for custom adaptations.

\subsubsection{Framework Execution}
Each TUS approach is executed via a dedicated module that serves as its entry point and defines a function that returns a tuple consisting of three components: 
\begin{itemize}
    \item a dictionary of mapping steps to their corresponding execution and undo logic,
    \item a dictionary for tracking the execution status of each step,
    \item a file path for persistently storing the status information
\end{itemize}
The mapping steps are flexible and can be linked to any custom implementation, including functions, shell scripts, or other executable routines tailored to individual approaches. These steps must be manually defined by the admin of the data lake or the approach.

%
%
%

At benchmark runtime, the defined steps are executed sequentially, typically including embedding, indexing, and querying. If a specific step fails or needs to be rerun, TUSBench first invokes the corresponding undo function to revert its effects before re-execution, therefore, ensuring its consistent state. Upon successful completion of a step, its status is marked in the status dictionary, which is persisted to disk. During subsequent runs, this status file is used to identify completed steps, which are skipped unless explicitly re-executed, thus minimizing redundant computation. This mechanism also allows shared tasks, such as downloading model weights from HuggingFace or auxiliary resources, to persist across multiple executions of the same approach. Since input data adheres to a standardized format defined by the \textit{Intermediates \RNum{1}} interface, all approaches can be applied to any data lake without modifications on the approach side.

For each query table, the approach performs a top-$k$ search over the data lake and stores the resulting ranked list of predicted unionable tables for evaluation.

\subsubsection{Intermediates \RNum{2}}
The \textit{Intermediates \RNum{2}} interface standardizes the structure of prediction outputs so that all approaches can be evaluated in a consistent manner. Top-$k$ ranked lists are saved as a dictionary in a  \textit{prediction.pickle} file, mapping each query table to a ranked list of predicted unionable tables.

\begin{table*}[]
    \centering
    \caption{Specifications of Benchmark Data Lakes used in Evaluation.}
    \begin{tabular}{|c|c|c|c|c|c|c|c|c|c|c|}
    \hline
    \textbf{Data Lake} & \textbf{\#Queries} & \textbf{\#Tables} & \textbf{Avg. Hits} & \textbf{Avg. Cols} & \textbf{Avg. Rows} & \textbf{Size (MB)} & \textbf{String (\%)} & \textbf{Float (\%)} & \textbf{Int (\%)}  & \textbf{Bool (\%)} \\ \hline
    tus\_santos \cite{DBLP:journals/corr/abs-2307-04217} & 40 & 2040 & 13.69 & 10.16 & 4728.73 & 1500 & 85.46 & 6.45 & 8.09 & 0 \\ \hline
    wiki\_union \cite{DBLP:journals/corr/abs-2307-04217} & 7158 & 33594 & 11.78 & 2.59 & 51.13 & 142 & 60.06 & 25.22 & 14.72 & 0 \\ \hline
    ecb\_union \cite{DBLP:journals/corr/abs-2307-04217} & 508 & 3718 & 51.81 & 35.95 & 307.38 & 435 & 49.17 & 36.67 & 14.16 & 0 \\ \hline
    ugen\_v1 \cite{DBLP:journals/corr/abs-2308-03883} & 49 & 960 & 9.73 & 10.37 & 7.66 & 3.8 & 92.72 & 3.91 & 3.37 & 0 \\ \hline
    ugen\_v2 \cite{DBLP:journals/corr/abs-2308-03883} & 50 & 1000 & 10 & 13.36 & 18.74 & 8.1 & 82.83 & 5.50 & 11.67 & 0 \\ \hline
    webtable \cite{DBLP:journals/pvldb/DengCCYCYSWLCJZJZWYWT24} & 5476 & 34796 & 16.33 & 12.61 & 39.41 & 198 & 58.61 & 13.27 & 28.12 & 0 \\ \hline
    opendata \cite{DBLP:journals/pvldb/DengCCYCYSWLCJZJZWYWT24} & 3062 & 7925 & 18.95 & 19.07 & 73948.43 & 107000 & 52.40 & 24.30 & 23.04 & 0.26 \\ \hline
    \end{tabular}
    \label{datalakes}
\end{table*}

\subsubsection{Empirical Validation}
In the final evaluation phase, TUSBench computes performance metrics by comparing the top-$k$ results from the \textit{Intermediates \RNum{2}} interface with the ground truth from the \textit{Intermediates \RNum{1}} interface. This is achieved using TUSBench’s functionality to evaluate performance across varying values of $k$. Core metrics include, but are not limited to, Mean Average Precision and Average Recall, evaluated over a predefined list of $k$ values. These results are recorded in a resulting \textit{experiments.csv} file. In addition, the execution times for each processing step defined in the Framework Execution stage is tracked. By structuring the approach into finer-grained steps, users can obtain a more detailed breakdown of execution performance. All output files can optionally be saved to a shared repository.

\section{Experiments}
\label{sec:experiments}
We evaluate EasyTUS against state-of-the-art, publicly available baseline approaches, therefore excluding AutoTUS and TabSketchFM, using established benchmark data lakes within the TUSBench environment. To assess both time efficiency and accuracy, we measure execution time along with Mean Average Precision (MAP) and Average Recall (AR), two standard metrics commonly used in prior work to assess the ranking quality of top-$k$ Table Union Search results across multiple queries. To further highlight EasyTUS's generalization capabilities and avoid any risk of data contamination, the evaluation includes a synthetic, previously unseen data lake.

\subsection{Setup}
The data lakes in these experiments are sourced from previous studies and are listed in Table~\ref{datalakes}. They serve as the basis for benchmarking the current state-of-the-art. Both \textit{webtable} and \textit{opendata} are light-weight versions of the originally published data lakes \cite{DBLP:journals/pvldb/DengCCYCYSWLCJZJZWYWT24}, excluding the noise tables that are not unionable with any given query table.

To further assess robustness, each data lake is accompanied by an adapted variant in which all metadata has been removed to particularly demonstrate that EasyTUS not only generalizes across structured data lakes but also performs well in minimal-context scenarios, with no metadata. In these adapted variants, table names are replaced with unique hexadecimal identifiers, and column names are anonymized as $col_i$, where $0 \leq i < n$ and $n$ is the number of columns. Mean Average Precision (MAP) is computed uniformly using Equation~\ref{eq:map}, with a key modification: instead of a fixed $k$, we use $min(k, k^*)$, where $k^*$ is the number of unionable tables in the ground truth for a given query. This adjustment ensures that setting $k > k^*$ does not unfairly penalize approaches and eliminates the need to tailor $k$ per query. For $k$, values of $2^i$ with $0 \leq i \leq 6$ are selected, which on average cover the full range of target tables per query (Avg. Hits) reported in Table~\ref{datalakes}. 

\begin{equation}
\label{eq:map}
    MAP@k = \frac{1}{|Q|} \cdot \sum_{q \in Q} AP_q@k
\end{equation}
where
\begin{equation}
    AP_q@k = \frac{1}{min(k,k^*)} \cdot \sum_{i=1}^k P_{q,i} \cdot rel_{q,i}
\end{equation}
with $Q$ denoting the set of queries, $P_{q,i}$ the precision at position $i$ for query $q \in Q$, $k^*$ the number of unionable tables in the ground truth for query $q$, and $rel_{q,i} = 1$ if the table at rank $i$ is relevant, and $0$ otherwise.

Additionally, we evaluate the standard metric Recall using the Average Recall (AR), as defined in Equation~\ref{eq:recall}. 

\begin{equation}
    \label{eq:recall}
    AR@k = \frac{1}{|Q|} \cdot \sum_{q \in Q}\sum_{i=1}^k rel_{q,i}
\end{equation}

All experiments are conducted on a high-performance server equipped with 996GB RAM, an AMD EPYC 7662 64-core processor, Ubuntu 22.04.5 LTS, and a single NVIDIA A100 80GB GPU. Each approach is run using its default configuration, as specified in the original publications.

\begin{figure}[t!]
    \centering
    \begin{subfigure}[t]{0.5\linewidth}
        \centering
        \includegraphics[height=1.2in]{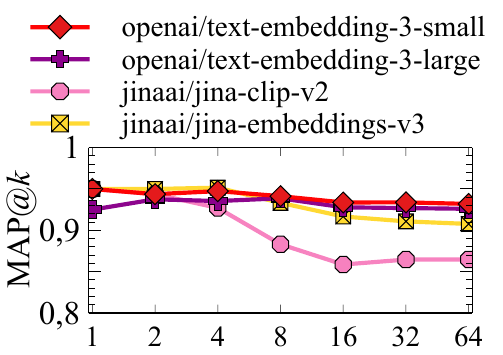}
        \caption{Different embedding models}
        \label{fig:model}
    \end{subfigure}%
    ~ 
    \begin{subfigure}[t]{0.5\linewidth}
        \centering
        \includegraphics[height=1.2in]{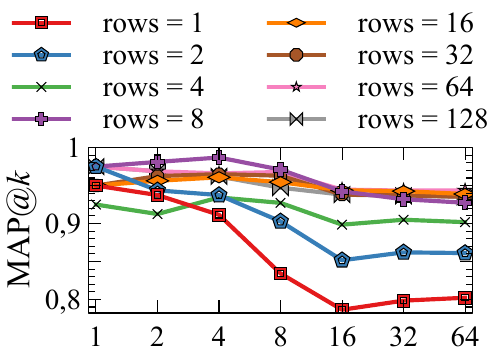}
        \caption{Different table-row sizes}
        \label{fig:rowcount}
    \end{subfigure}
    \caption{Mean Average Precision (MAP@$k$) on the tus\_santos data lake benchmark. Values of k are plotted on the x-axis. Similar trends are observed for other data lakes as well.}
\end{figure}

\subsection{LLM Selection}
 We maximize usability by minimizing manual parameter tuning while ensuring robust generalization. This also applies to selecting models for generating the embeddings. Since EasyTUS can support a wide range of LLM-based embedders, exhaustively evaluating all possible models is impractical. However, factors such as data privacy, resource availability, and model size play a crucial role in selecting the appropriate model. Therefore, we compare OpenAI’s cloud-based models \cite{openai_new_embedding_models_2024}, which offer large-scale capabilities, with JinaAI’s on-premise models \cite{DBLP:journals/corr/abs-2409-10173, DBLP:journals/corr/abs-2412-08802}, which prioritize data privacy. Figure~\ref{fig:model} shows MAP@$k$ scores for four embedding models. Due to space constraints, the tus\_santos data lake is shown as representative results, however, similar results were observed across other data lakes. Our findings show that OpenAI models yield slightly higher precision, though the difference is modest: JinaAI’s best model lags OpenAI’s lowest by only one percentage point and differences among OpenAI’s models are negligible. These findings suggest JinaAI is a practical alternative when data privacy outweighs slightly higher accuracy gains. For this reason, in the following experiments, we use both \textit{openai-text-embedding-3-large}, hereafter referred to as \textit{\mbox{ET-O}}, and \textit{jina-embeddings-v3}, referred to as \textit{\mbox{ET-J}}. This dual setup offers users flexibility in balancing precision and privacy, without compromising overall performance.


\subsection{Row Count}


The EasyTUS batching algorithm samples a set of random rows per table. To optimally balance diversity of data points against the risk of exceeding token limits, experiments were conducted on the tus\_santos data lake, again selected as a representative case, with similar results observed for the other data lakes. We evaluated system performance across row counts of $2^i$ with $0 \leq i \leq 7$, as shown in Figure~\ref{fig:rowcount}, using MAP@$k$ as the primary metric. Embeddings were generated using the \textit{openai-text-embedding-3-large} model as a representative example, with $n$ randomly selected rows (without repetition) per table; due to space constraints, only results for this model are shown, though all other tested models yielded comparable performance. The overall results indicate that row counts of 1–8 yield significantly lower MAP@$k$ scores, while performance stabilizes above 0.9 for row counts of 16 and higher. However, precision is not the sole performance criterion. Embedding latency and the frequency of table truncation due to token limitations are equally critical, as they influence the scalability and applicability of the approach to large-scale or high-complexity tabular data. For 128 rows, embedding initialization took 396 seconds, with 22.6\% of total 2,040 tables exceeding the 8,192-token context limit, thus, requiring row reduction. With 64 rows, the time dropped to 83 seconds (175 reductions), and at 32 rows, to just 66 seconds (52 reductions). These findings reveal that 16 to 32 rows strike the optimal balance: maintaining high accuracy (comparable to 64 and 128), while reducing computation time and minimizing token-related table reductions. Comparable results were obtained across the other data lakes and LLMs. As a result, we adopt 32 rows as the default. 

\begin{table*}[]
    \centering
    \caption{Execution Times in Seconds for \mbox{ET-O}, \mbox{ET-J}, D3L, Santos, and Starmie across Data Lakes.}
    \begin{tabular}{|c|c|c|c|c|c|c|c|c|c|c|}
    \hline
        \multirow{2}{*}{\textbf{Data Lake}} & \multicolumn{5}{c|}{\textbf{Offline Time (s)}} & \multicolumn{5}{c|}{\textbf{Online Time (s)}} \\ \cline{2-11}
        & ET-O & ET-J & D3L & Santos & Starmie & ET-O & ET-J & D3L & Santos & Starmie \\ \hline
         tus\_santos & 106.87 & 411.2 & 3726.82 & 39294.51 & 10484.51 & 4345.68 & 7.04 & 683.95 &  1220.81 & 3.57 \\
         wiki\_union & 469.99 & 1761.99 & 1049.48 & 21033.82 & 4587.16 & 693.77 & 106.3 & 410799.26 & 446.99 & 16.49 \\ 
         ecb\_union & 156.2 & 1738.23 & 1177.43 & 15419.13 & 4467.51 & 7155.68 & 13.51 & 248823.55 & 30477.76 & 479.52 \\
         ugen\_v1 & 29.55 & 104.24 & 186.04 & 3764.89 & 330.02 & 2177.16 & 13.65 & 611.48 & 86.28 & 5.4 \\
         ugen\_v2 & 33.43 & 164.64 & 190.79 & 3629.16 & 287.45 & 1870.99 & 6.56 & 161.44 & 72.85 & 3,02 \\
         opendata & 6279.12 & 4138.58 & 603670.9 & - & 573017.87 & 78.08 & 46.62 & 99238.98 & - & 1723,64 \\
         webtable & 472.03 & 1305.76 & 1783.4 & 33432.14 & 4875.06 & 34447.39 & 103.04 & 300455.58 & 28564.22 & 60.11 \\ \hline\hline
         \textbf{Average} & \textbf{1078.17} & 1374.95 & 87397.84 & 19428.94 & 85435.65 & 7252.68 & \textbf{42.39} & 182364.27 & 10144.82 & 327.39 \\ \hline 
    \end{tabular}
    \label{tab:efficiency}
\end{table*}

\subsection{Accuracy}
\begin{figure}[!th]
    \centering
    \includegraphics[width=\linewidth]{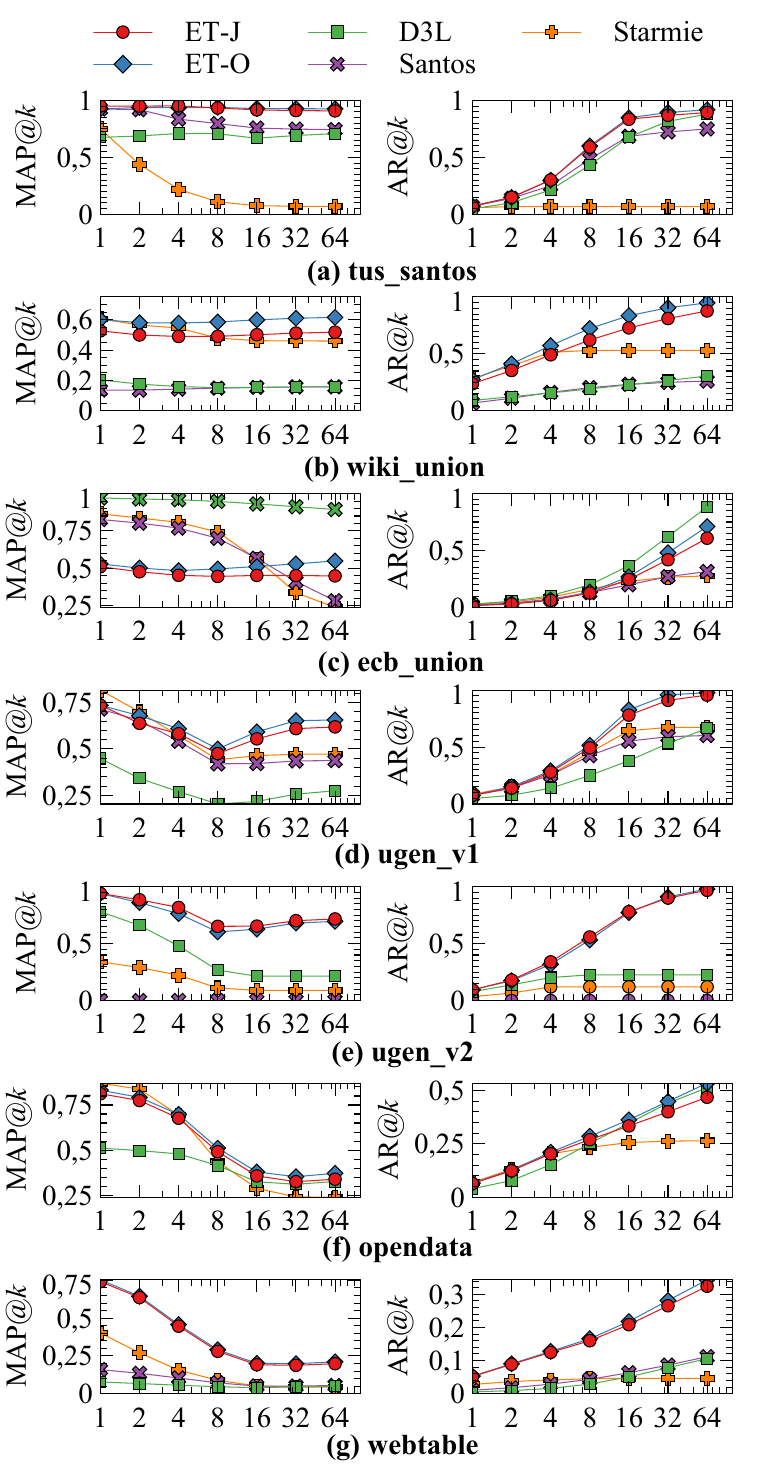}
    \caption{Comparison of EasyTUS (\mbox{ET-J} and \mbox{ET-O}) against the state-of-the-art in terms of Mean Average Precision (MAP@$k$) and Average Recall (AR@$k$) across selected benchmark data lakes with metadata. Values of $k$ are plotted on the x-axis.}
    \label{fig:with_metadata}
\end{figure}
\begin{figure}[!th]
    \centering
    \includegraphics[width=\linewidth]{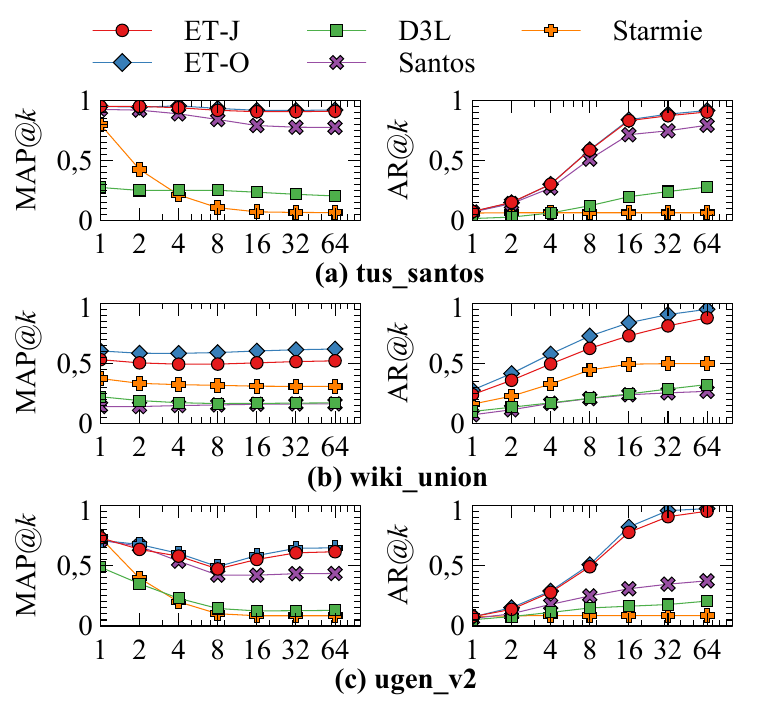}
    \caption{Comparison of EasyTUS (\mbox{ET-J} and \mbox{ET-O}) against the state-of-the-art in terms of Mean Average Precision (MAP@$k$) and Average Recall (AR@$k$) across selected benchmark data lakes without metadata. Values of $k$ are plotted on the x-axis.}
    \label{fig:without_metadata}
\end{figure}

We evaluate the accuracy of EasyTUS against the leading baseline Table Union Search systems Starmie, Santos, and D3L across all given data lakes. All experiments use a consistent configuration of 32 rows and the cosine similarity as distance metric. Figure~\ref{fig:with_metadata} presents the results across data lakes (a–g) with metadata included. To evaluate robustness under limited metadata conditions, each data lake was also prepared in a version with metadata removed, as shown in Figure~\ref{fig:without_metadata}. Due to space constraints, only a subset of those results is shown here for illustration. For each benchmark data lake, both MAP@$k$ and AR@$k$ are evaluated.

When metadata is available, the majority of approaches initially achieve high MAP scores. In many cases, across $k$ values and data lakes, the EasyTUS variants \mbox{ET-J} and \mbox{ET-O} achieve the highest MAP scores, generally performing on par with each other. A notable exception appears in the \textit{wiki\_union} data lake (sub-figure 1b), where \mbox{ET-O} performs better than \mbox{ET-J}. For Average Recall, \mbox{ET-J} and \mbox{ET-O} typically identify relevant tables more quickly than competing approaches. The improvement is computed as the average over all values of $k$ for each approach within a data lake, and then averaged across all data lakes. \mbox{ET-O} achieves a 34.33\% improvement and \mbox{ET-J} a 28.89\% improvement over the next-highest precision, obtained by D3L.
In the \textit{ecb\_union} data lake (Figure~\ref{fig:with_metadata}c), EasyTUS is outperformed by all other approaches for $k \lesssim 16$, with D3L achieving the highest MAP and AR before and beyond this point. This behavior is likely due to the composition of the data lake, which, as shown in Table~\ref{datalakes}, contains predominantly numerical values that are difficult to interpret without fine-tuning. Even with fine-tuning, composite search approaches such as D3L, which maintain a dedicated index for numerical values, have a clear advantage. In summary, EasyTUS outperforms all baseline systems across most benchmark data lakes, with the exception of cases where numerical values dominate over strings.

When metadata is removed, some approaches experience a significant drop in accuracy, e.g. D3L on tus\_santos and Starmie on wiki\_union. Interestingly, ugen\_v2 even shows improved accuracy for Starmie and Santos. Only EasyTUS demonstrates consistently robust performance across all data lakes including ecb\_union, regardless of the absence of metadata. This makes EasyTUS particularly well-suited for data lakes, as its performance is entirely independent of metadata.


\subsection{Time Efficiency}
Table~\ref{tab:efficiency} reports offline and online execution times, whose sum gives the total end-to-end runtime for each approach across all data lakes. We terminated the execution of Santos on opendata after 527 hours, as it was impossible to surpass the best state-of-the-art results. The results show that EasyTUS, particularly with the JinaAI model (\mbox{ET-J}), on average, outperforms most competing approaches. Speedup is computed by averaging execution times per approach within each data lake and then averaging across all lakes. {ET-O} achieves an offline speedup of 79.24× and \mbox{ET-J} 62.14× compared to the next-fastest approach, Starmie. For online execution, \mbox{ET-J} is 7.72× faster than Starmie. In contrast, \mbox{ET-O} suffers from network latency, making it 22.15× slower online, underscoring the trade-off between local LLMs and cloud-based models in terms of both accuracy and time efficiency. Overall, EasyTUS delivers the fastest average execution times across all evaluated approaches.

\subsection{Data Contamination}
\begin{figure}
    \centering
    \includegraphics[width=\linewidth]{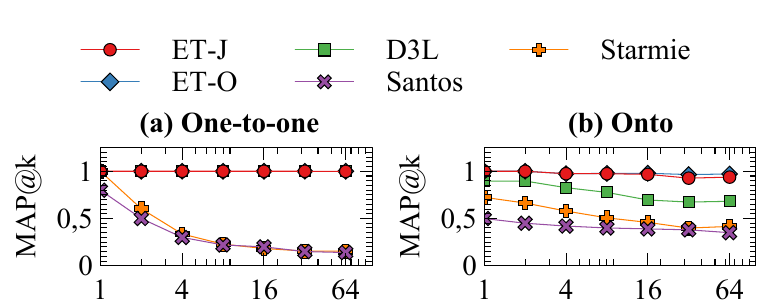}
    \caption{Comparison of Table Union Search approaches on an unseen synthetic data lake with (a) one-to-one and (b) onto mappings. Values of $k$ are plotted on the x-axis.}
    \label{fig:private}
\end{figure}

Data contamination refers to the risk of evaluating LLMs on benchmark data lakes that may have been included in their training data, thereby undermining the validity of experimental results \cite{DBLP:journals/corr/abs-2406-04244}. To assess EasyTUS’s generalization capabilities and its robustness to truly unseen data, we constructed a synthetic data lake using only tables from open data sources that were published after the release dates of the LLMs employed. This ensures the exclusion of any overlap with the pretraining data of respective LLMs. To support well-defined union operations, we created two distinct data lakes following \textit{one-to-one} and \textit{onto} mapping strategies. For both settings, each source table was randomly partitioned horizontally: the initial partition served as the query table, while the remaining partitions were assigned to the data lake. 
We enforced strict domain separation to prevent semantic overlap between unrelated tables. To better reflect real-world conditions, we injected noise tables, i.e. tables that have no unionable counterparts. For the onto dataset variant, we further randomly removed columns from the target tables. As illustrated in Figure~\ref{fig:private}, EasyTUS achieves performance on par with or better than state-of-the-art in the (a) one-to-one setting, and significantly outperforms them in the more challenging (b) onto scenario, as measured by Mean Average Precision. These results underscore the robustness and domain-agnostic generalization of EasyTUS.

\section{Conclusion}
In this paper, we introduced \textit{EasyTUS}, a faster and more accurate LLM-based comprehensive framework to the Table Union Search problem. By leveraging off-the-shelf models pre-trained on broad and diverse corpora, EasyTUS demonstrates superior performance without the need for fine-tuning and offers broad applicability across different types of data lakes. Importantly, it operates solely on raw data, while ignoring the metadata to avoid the pitfalls of ambiguity, inaccuracy, or incompleteness. The framework also includes \textit{TUSBench}, a standardized and modular testbed for evaluating the existing and future Table Union Search approaches. TUSBench enables independent and reproducible benchmarking across different approaches, data lakes, and evaluation metrics. Our experiments show that EasyTUS achieves, on average, up to 34.3\% higher Mean Average Precision, while being up to 79.2× faster in data preparation and up to 7.7× faster in query execution compared to the best prior approaches. Thanks to its modular architecture and support for seamlessly integrating newer, more efficient, and effective models, EasyTUS is well-positioned to maintain its performance advantage over other approaches without the need for expensive fine-tuning.

\bibliographystyle{IEEEtran}
\bibliography{paper}

\begin{thebibliography}{10}
\providecommand{\url}[1]{#1}
\csname url@samestyle\endcsname
\providecommand{\newblock}{\relax}
\providecommand{\bibinfo}[2]{#2}
\providecommand{\BIBentrySTDinterwordspacing}{\spaceskip=0pt\relax}
\providecommand{\BIBentryALTinterwordstretchfactor}{4}
\providecommand{\BIBentryALTinterwordspacing}{\spaceskip=\fontdimen2\font plus
\BIBentryALTinterwordstretchfactor\fontdimen3\font minus \fontdimen4\font\relax}
\providecommand{\BIBforeignlanguage}[2]{{%
\expandafter\ifx\csname l@#1\endcsname\relax
\typeout{** WARNING: IEEEtran.bst: No hyphenation pattern has been}%
\typeout{** loaded for the language `#1'. Using the pattern for}%
\typeout{** the default language instead.}%
\else
\language=\csname l@#1\endcsname
\fi
#2}}
\providecommand{\BIBdecl}{\relax}
\BIBdecl

\bibitem{DBLP:journals/tkde/HaiKQJ23}
\BIBentryALTinterwordspacing
R.~Hai, C.~Koutras, C.~Quix, and M.~Jarke, ``Data lakes: {A} survey of functions and systems,'' \emph{{IEEE} Trans. Knowl. Data Eng.}, vol.~35, no.~12, pp. 12\,571--12\,590, 2023. [Online]. Available: \url{https://doi.org/10.1109/TKDE.2023.3270101}
\BIBentrySTDinterwordspacing

\bibitem{DBLP:journals/pvldb/NargesianZMPA19}
F.~Nargesian, E.~Zhu, R.~J. Miller, K.~Q. Pu, and P.~C. Arocena, ``Data lake management: Challenges and opportunities,'' \emph{Proc. {VLDB} Endow.}, vol.~12, no.~12, pp. 1986--1989, 2019.

\bibitem{DBLP:journals/pvldb/NargesianZPM18}
F.~Nargesian, E.~Zhu, K.~Q. Pu, and R.~J. Miller, ``Table union search on open data,'' \emph{Proc. {VLDB} Endow.}, vol.~11, no.~7, pp. 813--825, 2018.

\bibitem{DBLP:conf/icde/BogatuFP020}
\BIBentryALTinterwordspacing
A.~Bogatu, A.~A.~A. Fernandes, N.~W. Paton, and N.~Konstantinou, ``Dataset discovery in data lakes,'' in \emph{36th {IEEE} International Conference on Data Engineering, {ICDE} 2020, Dallas, TX, USA, April 20-24, 2020}.\hskip 1em plus 0.5em minus 0.4em\relax {IEEE}, 2020, pp. 709--720. [Online]. Available: \url{https://doi.org/10.1109/ICDE48307.2020.00067}
\BIBentrySTDinterwordspacing

\bibitem{DBLP:journals/pacmmod/KhatiwadaFSCGMR23}
\BIBentryALTinterwordspacing
A.~Khatiwada, G.~Fan, R.~Shraga, Z.~Chen, W.~Gatterbauer, R.~J. Miller, and M.~Riedewald, ``{SANTOS:} relationship-based semantic table union search,'' \emph{Proc. {ACM} Manag. Data}, vol.~1, no.~1, pp. 9:1--9:25, 2023. [Online]. Available: \url{https://doi.org/10.1145/3588689}
\BIBentrySTDinterwordspacing

\bibitem{DBLP:journals/pvldb/FanWLZM23}
G.~Fan, J.~Wang, Y.~Li, D.~Zhang, and R.~J. Miller, ``Semantics-aware dataset discovery from data lakes with contextualized column-based representation learning,'' \emph{Proc. {VLDB} Endow.}, vol.~16, no.~7, pp. 1726--1739, 2023.

\bibitem{DBLP:conf/acl/HuWQLSFKKWY23}
\BIBentryALTinterwordspacing
X.~Hu, S.~Wang, X.~Qin, C.~Lei, Z.~Shen, C.~Faloutsos, A.~Katsifodimos, G.~Karypis, L.~Wen, and P.~S. Yu, ``Automatic table union search with tabular representation learning,'' in \emph{Findings of the Association for Computational Linguistics: {ACL} 2023, Toronto, Canada, July 9-14, 2023}, A.~Rogers, J.~L. Boyd{-}Graber, and N.~Okazaki, Eds.\hskip 1em plus 0.5em minus 0.4em\relax Association for Computational Linguistics, 2023, pp. 3786--3800. [Online]. Available: \url{https://doi.org/10.18653/v1/2023.findings-acl.233}
\BIBentrySTDinterwordspacing

\bibitem{DBLP:journals/corr/abs-2407-01619}
\BIBentryALTinterwordspacing
A.~Khatiwada, H.~Kokel, I.~Abdelaziz, S.~Chaudhury, J.~Dolby, O.~Hassanzadeh, Z.~Huang, T.~Pedapati, H.~Samulowitz, and K.~Srinivas, ``Tabsketchfm: Sketch-based tabular representation learning for data discovery over data lakes,'' \emph{CoRR}, vol. abs/2407.01619, 2024. [Online]. Available: \url{https://doi.org/10.48550/arXiv.2407.01619}
\BIBentrySTDinterwordspacing

\bibitem{DBLP:journals/corr/abs-2402-07927}
\BIBentryALTinterwordspacing
P.~Sahoo, A.~K. Singh, S.~Saha, V.~Jain, S.~Mondal, and A.~Chadha, ``A systematic survey of prompt engineering in large language models: Techniques and applications,'' \emph{CoRR}, vol. abs/2402.07927, 2024. [Online]. Available: \url{https://doi.org/10.48550/arXiv.2402.07927}
\BIBentrySTDinterwordspacing

\bibitem{DBLP:journals/pvldb/PaulsenGD23}
D.~Paulsen, Y.~Govind, and A.~Doan, ``Sparkly: {A} simple yet surprisingly strong {TF/IDF} blocker for entity matching,'' \emph{Proc. {VLDB} Endow.}, vol.~16, no.~6, pp. 1507--1519, 2023.

\bibitem{korade2024strengthening}
N.~B. Korade, M.~B. Salunke, A.~A. Bhosle, P.~B. Kumbharkar, G.~G. Asalkar, and R.~G. Khedkar, ``Strengthening sentence similarity identification through openai embeddings and deep learning.'' \emph{International Journal of Advanced Computer Science \& Applications}, 2024.

\bibitem{huggingface_2023}
{Hugging Face}, ``Hugging face: The ai community building the future,'' \url{https://huggingface.co}, 2023, accessed on August 22, 2025.

\bibitem{DBLP:journals/pami/MalkovY20}
\BIBentryALTinterwordspacing
Y.~A. Malkov and D.~A. Yashunin, ``Efficient and robust approximate nearest neighbor search using hierarchical navigable small world graphs,'' \emph{{IEEE} Trans. Pattern Anal. Mach. Intell.}, vol.~42, no.~4, pp. 824--836, 2020. [Online]. Available: \url{https://doi.org/10.1109/TPAMI.2018.2889473}
\BIBentrySTDinterwordspacing

\bibitem{DBLP:conf/www/SuchanekKW07}
\BIBentryALTinterwordspacing
F.~M. Suchanek, G.~Kasneci, and G.~Weikum, ``Yago: a core of semantic knowledge,'' in \emph{Proceedings of the 16th International Conference on World Wide Web, {WWW} 2007, Banff, Alberta, Canada, May 8-12, 2007}, C.~L. Williamson, M.~E. Zurko, P.~F. Patel{-}Schneider, and P.~J. Shenoy, Eds.\hskip 1em plus 0.5em minus 0.4em\relax {ACM}, 2007, pp. 697--706. [Online]. Available: \url{https://doi.org/10.1145/1242572.1242667}
\BIBentrySTDinterwordspacing

\bibitem{DBLP:journals/pvldb/DengCCYCYSWLCJZJZWYWT24}
Y.~Deng, C.~Chai, L.~Cao, Q.~Yuan, S.~Chen, Y.~Yu, Z.~Sun, J.~Wang, J.~Li, Z.~Cao, K.~Jin, C.~Zhang, Y.~Jiang, Y.~Zhang, Y.~Wang, Y.~Yuan, G.~Wang, and N.~Tang, ``Lakebench: {A} benchmark for discovering joinable and unionable tables in data lakes,'' \emph{Proc. {VLDB} Endow.}, vol.~17, no.~8, pp. 1925--1938, 2024.

\bibitem{DBLP:journals/corr/abs-2505-21329}
\BIBentryALTinterwordspacing
A.~Boutaleb, B.~Amann, H.~Naacke, and R.~Angarita, ``Something's fishy in the data lake: {A} critical re-evaluation of table union search benchmarks,'' \emph{CoRR}, vol. abs/2505.21329, 2025. [Online]. Available: \url{https://doi.org/10.48550/arXiv.2505.21329}
\BIBentrySTDinterwordspacing

\bibitem{DBLP:journals/corr/abs-2307-04217}
\BIBentryALTinterwordspacing
K.~Srinivas, J.~Dolby, I.~Abdelaziz, O.~Hassanzadeh, H.~Kokel, A.~Khatiwada, T.~Pedapati, S.~Chaudhury, and H.~Samulowitz, ``Lakebench: Benchmarks for data discovery over data lakes,'' \emph{CoRR}, vol. abs/2307.04217, 2023. [Online]. Available: \url{https://doi.org/10.48550/arXiv.2307.04217}
\BIBentrySTDinterwordspacing

\bibitem{DBLP:journals/corr/abs-2308-03883}
\BIBentryALTinterwordspacing
K.~Pal, A.~Khatiwada, R.~Shraga, and R.~J. Miller, ``Generative benchmark creation for table union search,'' \emph{CoRR}, vol. abs/2308.03883, 2023. [Online]. Available: \url{https://doi.org/10.48550/arXiv.2308.03883}
\BIBentrySTDinterwordspacing

\bibitem{openai_new_embedding_models_2024}
{OpenAI}, ``New embedding models and api updates,'' \url{https://openai.com/index/new-embedding-models-and-api-updates/}, Jan.~25 2024.

\bibitem{DBLP:journals/corr/abs-2409-10173}
\BIBentryALTinterwordspacing
S.~Sturua, I.~Mohr, M.~K. Akram, M.~G{\"{u}}nther, B.~Wang, M.~Krimmel, F.~Wang, G.~Mastrapas, A.~Koukounas, N.~Wang, and H.~Xiao, ``jina-embeddings-v3: Multilingual embeddings with task lora,'' \emph{CoRR}, vol. abs/2409.10173, 2024. [Online]. Available: \url{https://doi.org/10.48550/arXiv.2409.10173}
\BIBentrySTDinterwordspacing

\bibitem{DBLP:journals/corr/abs-2412-08802}
\BIBentryALTinterwordspacing
A.~Koukounas, G.~Mastrapas, B.~Wang, M.~K. Akram, S.~Eslami, M.~G{\"{u}}nther, I.~Mohr, S.~Sturua, S.~Martens, N.~Wang, and H.~Xiao, ``jina-clip-v2: Multilingual multimodal embeddings for text and images,'' \emph{CoRR}, vol. abs/2412.08802, 2024. [Online]. Available: \url{https://doi.org/10.48550/arXiv.2412.08802}
\BIBentrySTDinterwordspacing

\bibitem{DBLP:journals/corr/abs-2406-04244}
\BIBentryALTinterwordspacing
C.~Xu, S.~Guan, D.~Greene, and M.~T. Kechadi, ``Benchmark data contamination of large language models: {A} survey,'' \emph{CoRR}, vol. abs/2406.04244, 2024. [Online]. Available: \url{https://doi.org/10.48550/arXiv.2406.04244}
\BIBentrySTDinterwordspacing

\end{thebibliography}

\end{document}